\documentclass[preprint2]{aastex}
\usepackage[T1]{fontenc}
\usepackage{graphics}

\makeatletter

\providecommand{\LyX}{L\kern-.1667em\lower.25em\hbox{Y}\kern-.125emX\@}
\let\SF@@footnote\footnote
\def\footnote{\ifx\protect\@typeset@protect
    \expandafter\SF@@footnote
  \else
    \expandafter\SF@gobble@opt
  \fi
}
\expandafter\def\csname SF@gobble@opt \endcsname{\@ifnextchar[
  \SF@gobble@twobracket
  \@gobble
}
\edef\SF@gobble@opt{\noexpand\protect
  \expandafter\noexpand\csname SF@gobble@opt \endcsname}
\def\SF@gobble@twobracket[#1]#2{}

\usepackage[first]{draftcopy}
\draftcopyName{Accepted}{180}

\makeatother

\begin{document}

\title{Ages of late spectral type Vega--like stars}

\author{Inseok Song, J.-P. Caillault,}

\affil{Department of Physics and Astronomy, University of Georgia, Athens GA 30602-2451
USA}

\email{song@physast.uga.edu, jpc@akbar.physast.uga.edu}

\author{David Barrado y Navascu\'{e}s,}

\affil{Max-Planck Institut für Astronomie, Königstuhl 17, Heidelberg, D-69117 Germany}

\email{barrado@mpia-hd.mpg.de}

\author{John R. Stauffer}

\affil{Harvard Smithsonian Center for Astrophysics, 60 Garden St., Cambridge, MA 02138
USA}

\email{stauffer@amber.harvard.edu}

\author{Sofia Randich}

\affil{Osservatorio Astrofisico di Arcetri, Largo E. Fermi 5, I-50125, Firenze Italy}

\email{randich@arcetri.astro.it}

\begin{abstract}
We have estimated the ages of eight late--type Vega--like stars by using standard
age dating methods for single late--type stars, e.g., location on the color
magnitude diagram, Li~6708 \AA{} absorption, CaII~H\&K emission, X--ray luminosity,
and stellar kinematic population. With the exception of the very unusual pre-main
sequence star system HD~98800, all the late--type Vega-like stars are the same
age as the Hyades cluster (600--800 Myr) or older.
\end{abstract}


\section{Introduction}

Vega--like stars are stars with IR excess emission due to optically thin dust
disks with little or no gas \citep{BP93,LBA99}. It is important to know the
ages of Vega--like stars because they may provide crucial information about
extra--solar planetary systems. Given the existence of a complete, flux--limited
IR survey and a 100\% success rate for identifying stars with IR excesses due
to circumstellar disks, then if all identified Vega--like stars were young,
one could conclude that the timescale for the disks to ``go away'' is short.
Furthermore, if the fraction of the number of Vega--like stars to the total
number of stars (\( N_{V}/N_{T} \)) is similar to the fraction of the average
ages of Vega--like stars to the average ages of milky way disk stars (\( T_{V}/T_{mw} \)),
then one could also conclude that most stars are born with such disks. Hence,
if such Vega disks are signposts of planet formation, then one could infer that
most stars have planetary systems. On the other hand, if Vega--like stars have
a wide range of ages and \( N_{V}/N_{T}\ll T_{V}/T_{mw} \), then one could
plausibly conclude that disk formation -- and hence possibly planet formation
-- is not a common process, at least at the level of sensitivity of the IR survey
used to identify the Vega--like stars.

There have been many efforts to determine the ages of Vega--like stars. \citet{BP93}
estimated the age of \( \beta  \)~\-Pictoris as 100 Myr whereas \citet{Jura93}
claimed that \( \beta  \)~\-Pictoris is extremely young, \( \sim 2-3 \) Myr.
They interpreted the same data in opposite ways, the latter group claiming that
\( \beta  \)~\-Pictoris is approaching the zero age main sequence (ZAMS) and
the former group believing that \( \beta  \)~\-Pictoris is an old main sequence
(MS) star. More recent studies of \( \beta  \)~Pictoris support its young age \citep{Jura98,BSSC}.
Another conspicuous Vega-like star, HR~4796A, is believed to be even younger
than \( \beta  \)~\-Pictoris \citep{Jura93,Stauffer95} while Fomalhaut and
Vega are estimated to be much older, \( \sim 200 \) Myr \citep{BP93,Barrado97,David98}.

In order to increase the number of age estimates of Vega--like stars, we have
compiled a list of 361 Vega--like stars from the literature and have searched
for late--type binary companions in order to follow the techniques of \citet{Stauffer95},
who determined the age of HR 4796A by using the properties of its late--type
companion star HR 4796B. The complete results of our survey can be found in
\citet{myPhD}. However, there are eight late--type Vega--like stars in our list
for which we can estimate ages directly by using standard age dating methods
for single late--type stars, methods such as their location on the CMD, Li 6708\AA\
absorption, X--ray luminosity, rotation, CaII H\&K emission, etc. 

In section 2, we present the late--type Vega--like stars and their age related
data. The age determinations are discussed in section 3, and a conclusion is
provided in section 4.

\section{Data}

In Table 1, we list the eight late--type (\( B-V>0.65 \)) Vega--like stars
along with the data to be used for age estimations. All spectral types are taken
from SIMBAD and X--ray luminosities are from the Rosat All Sky Survey data \citep{Hunsch98,Hunsch99}.
The fractional IR luminosity \( f \) (in the 6th column of Table 1) is defined
as \( f\equiv L_{IR}/L_{*} \). IRAS PSC2/FSC fluxes were used to calculate
the \( f \) values. The values of the logarithm of the ratio between CaII~H\&K
fluxes and stellar bolometric luminosity, \( \log R^{'}(HK) \), are from \citet{Henry}
and the sources of the other parameters are indicated by footnotes to Table
\ref{table1}. 

HD~98800 is a quadruple system consisting of two spectroscopic binaries \citep{Torres}.
Since the \( L_{x} \) flux includes the emission from all four components,
we divided the published value by four to take the multiplicity into account.
The Li abundance for this system is high, \( \log N(\mathrm{Li})= \)3.1, 2.3,
and 3 for components Aa, Ba, and Bb, respectively \citep{Soderblom98}, and the
values for each component were used instead of the value from the whole system,
when possible. \( M_{v} \) values for each component were taken from \citet{Soderblom98},
and, according to \citet{Gehrz99}, component B emits most of the IR excess emission.

\section{Age estimations}

\subsection{Color--Magnitude Diagram (CMD) location}

The eight late--type Vega--like stars are plotted in Figure 1 with theoretical
solar abundance PMS evolutionary tracks from \citet{DM97}. We have used a ``tuned''
temperature scale to convert \( T_{eff} \) to \( (V-I)_{c} \) \citep{Stauffer95},
which locates the 100 Myr isochrone onto the locus of Pleiades stars. Except
for the HD~98800 system, we used Hipparcos \( (V-I)_{c} \) colors to plot the
CMD. With the exception of only HD~73752 and HD~98800, all the Vega--like stars
are located on or close to the ZAMS which makes it difficult to get an age estimation
with precision smaller than the MS lifetime of the late--type stars. A star
whose position in the CMD is above the ZAMS can either be in the PMS stage or
in the post ZAMS evolutionary stage. We need additional information to determine
the actual ages of these stars. As argued below, HD~73752 is probably in the
post MS evolutionary stage whereas HD~98800 is almost certainly a PMS star.
Except for HD~73752 and HD~98800, we are only able to conclude that the stars
in our sample are older than \( \sim 30 \) Myr based solely in their locations
in the CMD.

\subsection{Lithium abundance}

Four stars (\( \epsilon  \) Eri, HD~73752, HD~98800, and HD~221354) have had
their Li 6708\AA{} absorption strengths measured. Those values and the Li data
for the Pleiades, the Hyades, and the M34 cluster members are plotted in Figure
2. It appears that HD~98800 is younger than the Pleiades cluster and that HD~73752,
\( \epsilon  \) Eri, and HD~221352 are about the same age as the Hyades cluster
(\( 600-800\, \mathrm{Myr} \)).

\subsection{X-ray emission}

HD~98800 is again located above the Pleiades cluster in a plot of \( L_{x} \)
versus \( (B-V) \) (Fig. 3). HD~73752, HD~67199, and HD~69830 have X--ray luminosities
much lower than those of Hyades stars with similar \( (B-V) \) colors, so we
assign older ages to those stars. \( \epsilon  \) Eri, HD~53143, and HD~128400,
however, show about the same X--ray activity as Hyades stars so their ages are
probably similar.

\subsection{CaII H\&K}

By using the chromospheric activity -- age relation of \citet{Donahue}: 
\[
\log t=10.725-1.334R_{5}+0.4085R^{2}_{5}-0.0522R_{5}^{3},\]
 where \( t \) is the age in years, and \( R_{5} \) is defined as \( R_{HK}^{'}\times 10^{5} \),
we can obtain another estimate of the ages of \( \epsilon  \) Eri, HD~53143,
HD~67199, and HD~128400. These ages agree with the age estimations from the
previous sections. The age uncertainty can be obtained from the measurement
error of \( \log R^{'}(HK) \), \( \Delta \log R^{'}(HK)=\pm 0.052 \) \( (\Delta \tau \sim 200\, \mathrm{Myr}) \) \citep{Henry}.
We adopted these ages as the final age estimations for the stars with CaII H\&K
measurements because this method is the only one that provides a quantitative
estimate. Age estimation from the CMD can also be fairly accurate, but only
when a star is in its PMS phase. Note that the calibration of this age formula
is presumably tied to an age scale where the Pleiades is 75 Myr and \( \alpha  \)
Persei is 50 Myr old. If the new lithium depletion ages for these clusters (125
and 85 Myr, respectively -- \citeauthor{Stauffer99} \emph{et al.} {[}\citeyear{Stauffer98}, \citeyear{Stauffer99}{]})
are correct, then the CaII H\&K ages should be increased by \( \sim 50\% \).

\subsection{Kinematics}

All eight stars appear in the Hipparcos catalog but only six stars have radial
velocity measurements (from \citealt{Duflot} and references therein). We have
used these radial velocities and parallaxes, along with B1950 coordinates and
proper motions, to calculate \( (U,V,W) \) galactic velocity components following
\citet{JohnsonUVW}. The kinematic population of the Vega--like stars is determined
by using the population criteria from \citet{Leggett} - young disk stars are
those with \( -20<U<50 \) and \( -30<V<0, \) while old disk stars have an
eccentricity less than 0.5 in the \( UV \) plane, but lie outside the young
disk ellipsoid. Of the six Vega--like stars with \( (U,V,W) \) velocities,
only HD~98800 is located inside of the young disk star regime (Fig. 4).

\subsection{Other criteria}

Since all Vega-like stars with \( v\sin i \) measurements in this study are
slow rotators, we can not assign any meaningful ages from this method. Metallicities
of the Vega-like stars with {[}Fe/H{]} measurements are very close to the solar
value. Thus, the metallicity data cannot constrain ages of these stars, either.

\subsection{Final estimated ages}

All age estimations are summarized in Table~\ref{table2}. Based on its X--ray
luminosity, Li abundance, kinematics, and location on the CMD, we are confident
that HD~98800 is a PMS star, with a probable age of \( \sim 7\, \mathrm{Myr} \),
and almost certainly lying between 2 and 12 Myr old which is in good agreement
with \citeauthor{Soderblom98}'s \citeyear{Soderblom98} estimate of \( 10^{\, +10}_{\, -5} \)
Myr. Based on a similar set of evidence, HD~73752 on the other hand is almost
certainly a post MS star whose IR excess emission mechanism may be different
from that of the younger Vega--like stars. The six other stars in our study
lie close to the ZAMS in the CMD and seem to have ages \( \sim 1\, \mathrm{Gyr} \).
For HD~69830 and HD~73752, we assigned maximum ages of 2 Gyr because they have
about the same X-ray activity as HD~67199 which has an age determination from
CaII H\&K. However, the ages for these stars could be as small as the age of
the Hyades cluster (but not smaller than that). Therefore, we assign ages in
the range of \( 600-2000\, \mathrm{Myr} \) for these two stars. HD~221354 does
not have a good age estimation, but from the fact that it is on the MS and that
it belongs kinematically to the old disk population stars, we assign an age
of \( 1.0\pm 0.5 \) Gyr. 

\citet{Habing99} performed a study very similar to ours but mainly of early--type
stars. They found that most Vega--disks disappear very sharply around an age
of 400 Myr. \citet{uvbyBpaper} have also found that early--type Vega--like stars
are systematically young (\( <400 \) Myr) by using Str\"{o}mgren \emph{uvby\( \beta  \)}
photometry.

\subsection{K stars with upper limit in \protect\( f\protect \)}

As seen from Table~\ref{table2}, there are no Vega-like stars with ages in
the range between a few tens and a few hundreds of Myr in the spectral type
range in which we are interested. To have more data in that age range, we constructed
a list of young K stars based on their strong Li absorption or strong X-ray
emission \citep{Fisher,Jeff}. We collected age related data for these stars
from which we estimated their ages (see \citealt{myPhD}). Some of these stars
were detected with IRAS. Upper limits to \( f \) for these stars were calculated
by assuming that each system has a dust disk with temperature of 100 K and that
the system was barely undetected at 60 \( \mu m \). These stars are plotted
as open downward pointing triangles in Figure~\ref{f_age}, a plot of \( f \)
versus age for early--type and late--type Vega--like stars with good age estimates.
Two stars in the list of young K stars, Gl~150 and HD~68586, lie above the ZAMS
in a CMD, but, based on the other data available for these stars, we find that
they must be post MS stars; as a result, they are not plotted in Figure~\ref{f_age}.

We find Figure~\ref{f_age} both illuminating and puzzling. For the early--type
stars, the correlation of age and IR excess plus the generally young ages for
these prototypical objects suggest that many A stars form with disks and the
disk excesses decrease with age due to some evolutionary process \citep{BSSC}.
For the late--type Vega--like stars, the much older derived ages at least suggests
that the timescale for disk evolution is longer.

\section{Conclusion}

We have estimated ages of eight late-type Vega-like stars by using standard
age dating methods for a late-type star. Except for the PMS star HD~98800, all
Vega-like stars in this study are the same age as the Hyades cluster or older. 

Because we have only one late--type Vega--like star with an age less than \( \sim 500 \)
Myr, we cannot draw any strong conclusions about the evolutionary timescale
for disk evolution for K dwarfs. It would be very useful to obtain mid--IR data
(with the very sensitive SIRTF, for example) for the stars in the list of young
K stars. It would also be useful to obtain more complete age indicator data
for this sample in order to make the age estimation more precise.

\acknowledgements{IS and JPC acknowledge the support of NASA through grant NAG5--6902.}

\bibliographystyle{apjl}
\bibliography{Vegas}

\newpage
\onecolumn

\begin{table}

\caption{\label{table1}Late-type Vega-like stars and their age related data.}
{\noindent \centering \begin{tabular}{ccccccccc}
\hline 
{\scriptsize HD \#}&
{\scriptsize Other Name}&
{\scriptsize Sp. Type}&
{\scriptsize \( \mathrm{M}_{\mathrm{v}} \)}&
{\scriptsize \( (B-V) \)}&
{\scriptsize \( f(L_{IR}/L_{*})\times 10^{3} \)}&
{\scriptsize \( \log L_{x}(ergs/s) \)}&
{\scriptsize \( \log R^{'}(HK) \)}&
{\scriptsize \( \log N(\mathrm{Li}) \)}\\
\hline 
{\scriptsize 22049}&
{\scriptsize \( \epsilon  \) Eri}&
{\scriptsize K2V}&
{\scriptsize \( 6.18\pm 0.11 \)}&
{\scriptsize \( 0.88 \)}&
{\scriptsize 0.08}&
{\scriptsize 28.32}&
{\scriptsize -4.47}&
{\scriptsize 0.25\( ^{a} \)}\\
{\scriptsize 53143}&
{\scriptsize GL 260}&
{\scriptsize K1V}&
{\scriptsize \( 5.49\pm 0.14 \)}&
{\scriptsize \( 0.81 \)}&
{\scriptsize 0.18}&
{\scriptsize 28.69}&
{\scriptsize -4.52}&
{\scriptsize --}\\
{\scriptsize 67199}&
{\scriptsize --}&
{\scriptsize K1V}&
{\scriptsize \( 5.99\pm 0.06 \)}&
{\scriptsize \( 1.0 \)}&
{\scriptsize 0.24}&
{\scriptsize 27.87}&
{\scriptsize -4.72}&
{\scriptsize --}\\
{\scriptsize 69830}&
{\scriptsize GL 302}&
{\scriptsize K0V}&
{\scriptsize \( 5.45\pm 0.05 \)}&
{\scriptsize \( 0.76 \)}&
{\scriptsize 0.61}&
{\scriptsize 27.48}&
{\scriptsize --}&
{\scriptsize -- }\\
{\scriptsize 73752}&
{\scriptsize HR 3430}&
{\scriptsize G3/5V}&
{\scriptsize \( 3.55\pm 0.11 \)}&
{\scriptsize \( 0.73 \)}&
{\scriptsize 0.03}&
{\scriptsize 27.98}&
{\scriptsize --}&
{\scriptsize 1.3\( ^{b} \)}\\
{\scriptsize 98800Aa}&
{\scriptsize GL 2084}&
{\scriptsize K5V}&
{\scriptsize \( 6.06 \)}&
{\scriptsize \( 1.17 \)}&
{\scriptsize --}&
{\scriptsize 30.17}&
{\scriptsize --}&
{\scriptsize 3.1\( ^{c} \)}\\
\multicolumn{1}{r}{{\scriptsize Ba}}&
{\scriptsize --}&
{\scriptsize K7V}&
{\scriptsize \( 6.79 \)}&
{\scriptsize \( 1.37 \)}&
{\scriptsize 230}&
{\scriptsize 30.17}&
{\scriptsize --}&
{\scriptsize 2.3\( ^{c} \)}\\
\multicolumn{1}{r}{{\scriptsize Bb}}&
{\scriptsize --}&
{\scriptsize M1V}&
{\scriptsize \( 8.5: \)}&
{\scriptsize \( 1.41 \)}&
{\scriptsize --}&
{\scriptsize 30.17}&
{\scriptsize --}&
{\scriptsize 3:\( ^{c} \)}\\
{\scriptsize 128400}&
{\scriptsize --}&
{\scriptsize G5V}&
{\scriptsize \( 5.19\pm 0.07 \)}&
{\scriptsize \( 0.67 \)}&
{\scriptsize 3.2}&
{\scriptsize 28.59}&
{\scriptsize -4.56}&
{\scriptsize --}\\
{\scriptsize 221354}&
{\scriptsize GL 895.4}&
{\scriptsize K2V}&
{\scriptsize \( 5.63\pm 0.15 \)}&
{\scriptsize \( 0.84 \)}&
{\scriptsize 1.0}&
{\scriptsize --}&
{\scriptsize --}&
{\scriptsize 0.158\( ^{d} \)}\\
\hline 
\end{tabular}\scriptsize \par}
\( ^{a} \) \citet{Mallik} \( ^{b} \) \citet{Lebre} \( ^{c} \) \citet{Soderblom93}
and see section 2 \( ^{d} \) \citet{Fisher}
\end{table}

\begin{table}

\caption{\label{table2}Age estimations for the late--type Vega--like stars.}
{\centering \begin{tabular}{ccccccc}
\hline 
Name&
\multicolumn{5}{c}{age dating methods}&
Final age\\
\cline{2-6} 
&
\( L_{\mathrm{x}} \)&
CaII H\&K&
Li&
kinematics&
CMD&
(Myr)\\
\hline 
\( \epsilon  \) Eri&
\( \sim \mathrm{H} \)&
730&
\( \sim \mathrm{H} \)&
OD&
MS&
\( 730\pm 200 \)\\
HD 53143&
\( \sim \mathrm{H} \)&
965&
--&
OD&
MS&
\( 965\pm 200 \)\\
HD 67199&
\( >\mathrm{H} \)&
2000&
--&
--&
\( > \)MS&
\( 2000\pm 200 \)\\
HD 69830&
\( >\mathrm{H} \)&
--&
--&
OD&
MS&
\( 600-2000 \)\\
HD 73752&
\( >\mathrm{H} \)&
--&
\( \sim \mathrm{H} \)&
OD&
\( > \)MS&
\( 600-2000 \)\\
HD 98800&
\( <\mathrm{P} \) &
--&
\( <\mathrm{P} \)&
YD&
\( 7\pm 5 \)&
\( 7\pm 5 \)\\
HD 128400&
\( \sim \mathrm{H} \)&
1100&
--&
--&
MS&
\( 1100\pm 600 \)\\
HD 221354&
--&
--&
\( \succeq \mathrm{H} \)&
OD&
MS&
\( 1500\pm 1000 \)\\
\hline 
\multicolumn{7}{l}{ P=Pleiades, H=Hyades, OD=old disk population, YD=young disk population}\\
\end{tabular}\par}\end{table}

\begin{figure}
{\par\centering \resizebox*{0.9\columnwidth}{!}{\rotatebox{270}{\includegraphics{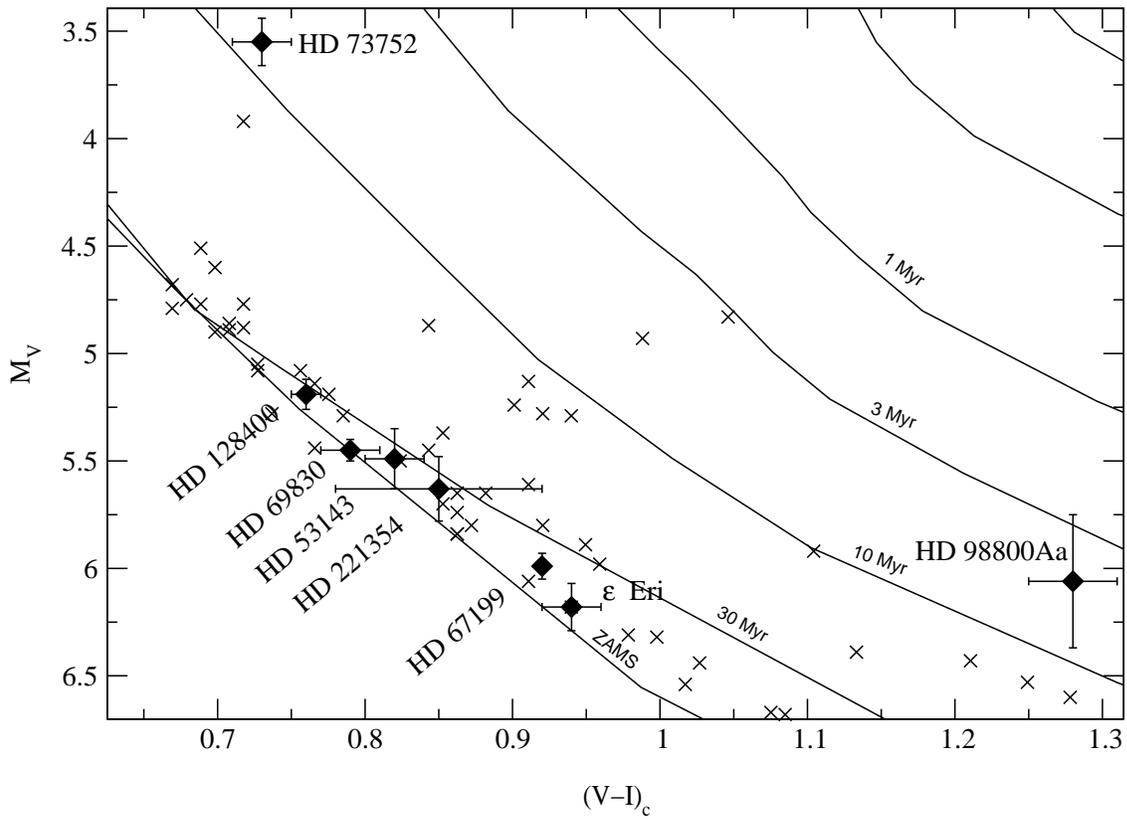}}} \par}

\caption{CMD of late--type Vega--like stars (filled diamonds) and the Pleiades stars
(crosses). Isochrones are from \citet{DM97}.}
\end{figure}
\begin{figure}
{\par\centering \resizebox*{0.9\columnwidth}{!}{\rotatebox{270}{\includegraphics{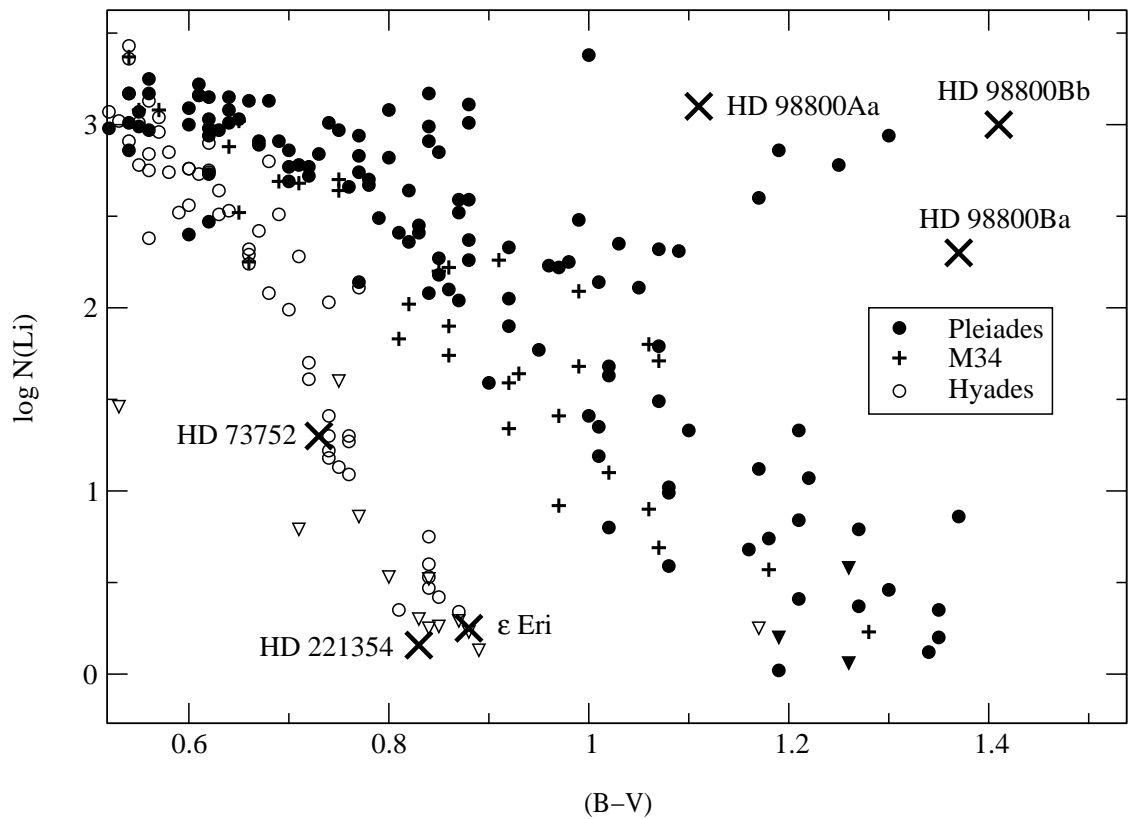}}} \par}

\caption{Li abundances of late--type Vega--like stars (large crosses) as well as members
of the Pleiades cluster (filled symbols, 125 Myr), M34 (plus signs, 250 Myr),
and the Hyades (open symbols, 600--800 Myr). Downward pointing triangles indicate
upper limits.}
\end{figure}
\begin{figure}
{\par\centering \resizebox*{0.9\columnwidth}{!}{\rotatebox{270}{\includegraphics{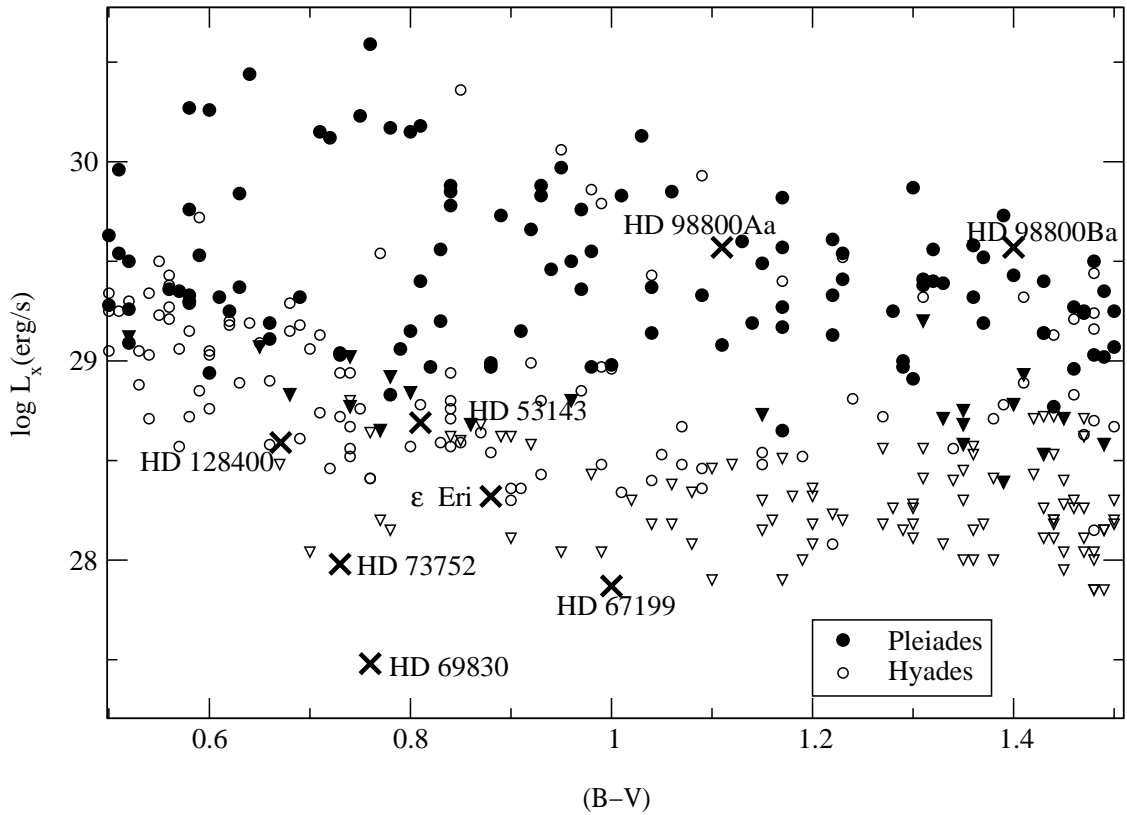}}} \par}

\caption{X--ray luminosities of late--type Vega--like stars, as well as those in the
Pleiades and the Hyades. All symbols have the same meaning as in Figure 2.}
\end{figure}
\begin{figure}
{\par\centering \resizebox*{0.9\columnwidth}{!}{\rotatebox{270}{\includegraphics{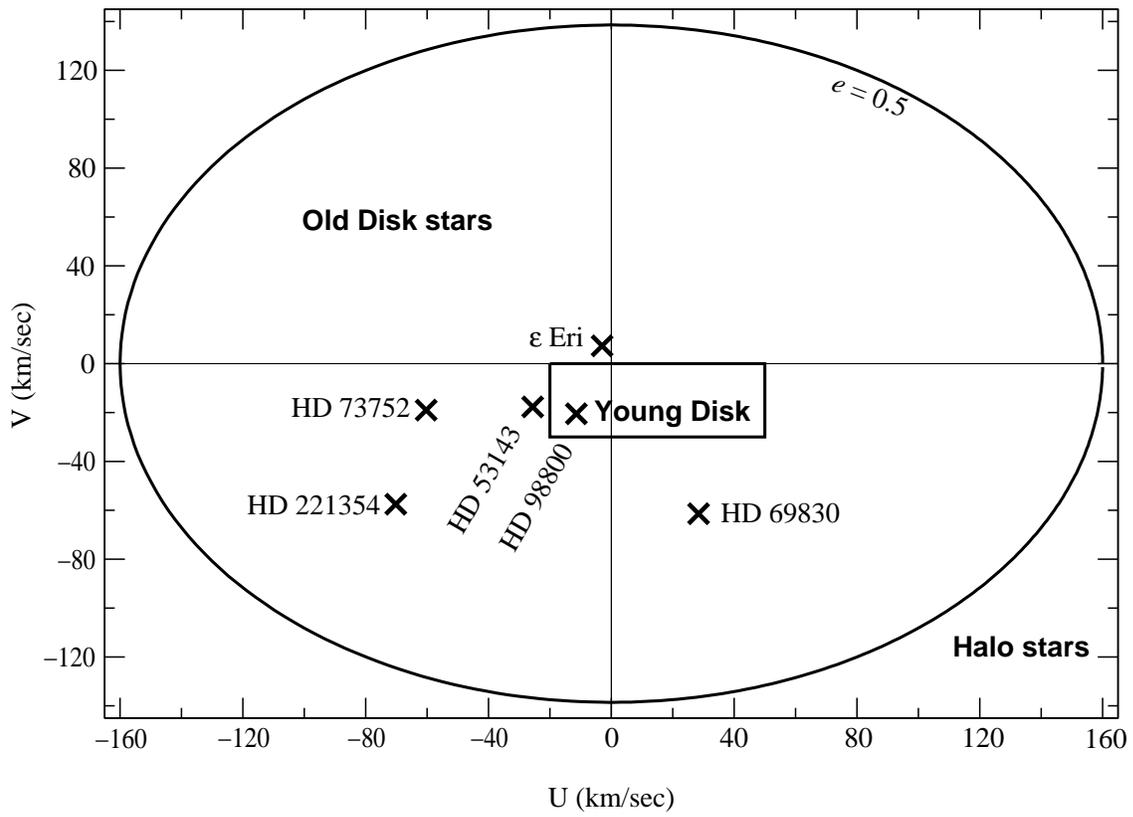}}} \par}

\caption{Kinematic population of late--type Vega--like stars. }
\end{figure}
\begin{figure}
{\par\centering \resizebox*{0.9\columnwidth}{!}{\rotatebox{270}{\includegraphics{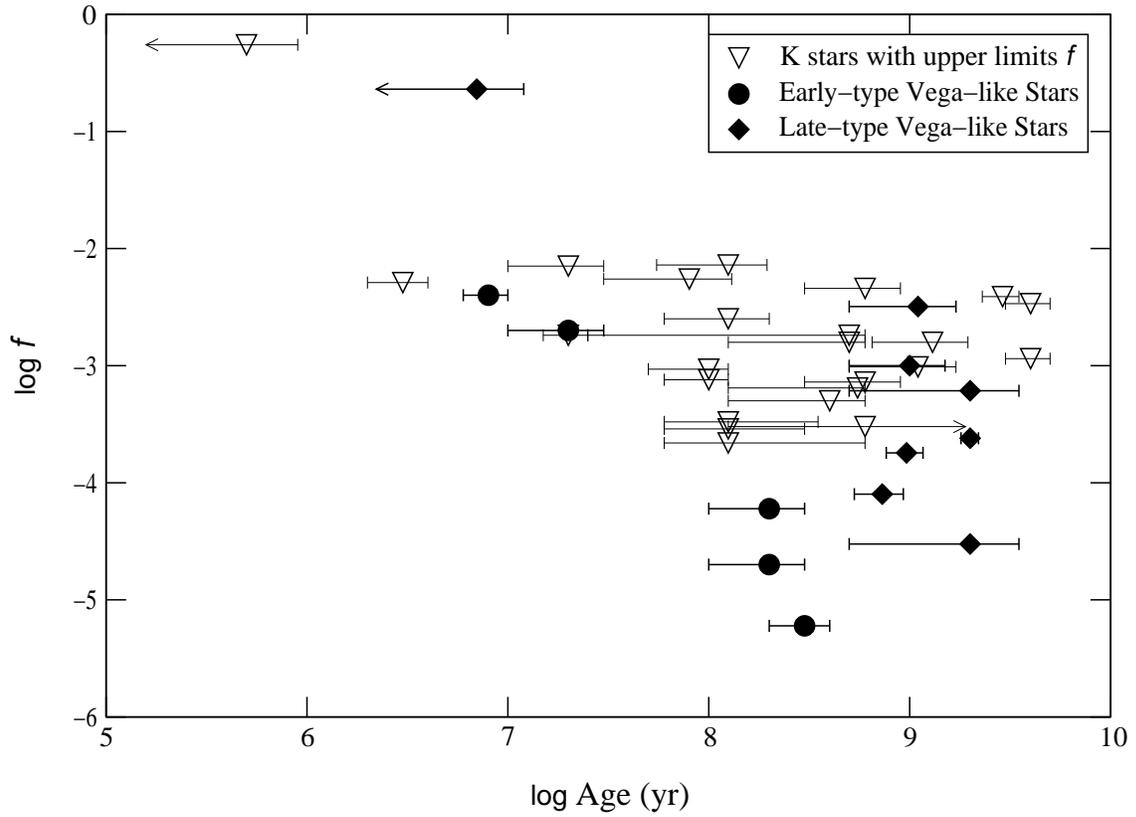}}} \par}

\caption{\label{f_age}Fractional infrared luminosity versus age. Early--type Vega--like
stars are plotted as solid circles and late--type Vega--like stars are plotted
as filled diamonds. Field K stars with IR excess upper limits are represented
with downward pointing open triangles.}
\end{figure}

\end{document}